\documentclass[pdftex,twocolumn,epjc3]{svjour3}          
\newcommand{\be}{\begin{equation}}\newcommand{\ee}{\end{equation}}
\newcommand{\bea}{\begin{eqnarray}}\newcommand{\eea}{\end{eqnarray}}
\newcommand{\brr}{\begin{array}}\newcommand{\err}{\end{array}}
\newcommand{\bit}{\begin{itemize}}\newcommand{\eit}{\end{itemize}}
\newcommand{\ben}{\begin{enumerate}}\newcommand{\een}{\end{enumerate}}

\newcommand{\ba}{\begin{array}}
\newcommand{\ea}{\end{array}}

\def\lf{\left}

\def\non{\nonumber}

\def\ri{\right}

\newcommand\myeq{\stackrel{\mathclap{\normalfont\mbox{\emph{a}}}}{\rightarrow}}

\def\la{\lambda}

\def\1{{_{1}}}\def\2{{_{2}}}

\def\noHe0{:\;\!\!\;\!\!:H_e(0):\;\!\!\;\!\!:}
\def\noHm0{:\;\!\!\;\!\!:H_\mu(0):\;\!\!\;\!\!:}

\def\lf{\left}

\def\non{\nonumber}

\def\ri{\right}

\def\la{\lambda}

\def\1{{_{1}}}\def\2{{_{2}}}

\RequirePackage[T1]{fontenc}

\smartqed  

\RequirePackage{graphicx}
\RequirePackage{mathptmx}      
\RequirePackage{flushend}
\RequirePackage[numbers,sort&compress]{natbib}
\RequirePackage[colorlinks,citecolor=blue,urlcolor=blue,linkcolor=blue]{hyperref}
\RequirePackage{amssymb}
\RequirePackage{graphics}
\RequirePackage{amsmath}
\RequirePackage{amsfonts}
\RequirePackage{bm}
\RequirePackage[]{latexsym}
\RequirePackage{epsfig}
\usepackage{mathtools,amssymb,lipsum}

\usepackage{cuted}
\usepackage{flushend}

\journalname{Eur. Phys. J. C}

\begin{document}

\title{On the $\beta$-decay of the accelerated proton and neutrino oscillations: a three-flavor description with CP
violation}

\author{Massimo Blasone\thanksref{e1,addr1,addr2}
      \and 
      Gaetano Lambiase\thanksref{e2,addr1,addr2}
      \and 
      Giuseppe Gaetano Luciano\thanksref{e3,addr1,addr2}
      \and
      Luciano Petruzziello\thanksref{e4,addr1, addr2}
}

\thankstext{e1}{blasone@sa.infn.it}
\thankstext{e2}{lambiase@sa.infn.it}
\thankstext{e3}{gluciano@sa.infn.it}
\thankstext{e4}{lpetruzziello@na.infn.it}

\institute{Dipartimento di Fisica, Universit\`a di Salerno, Via Giovanni Paolo II, 132 I-84084 Fisciano (SA), Italy. \label{addr1}
          \and INFN, Sezione di Napoli, Gruppo collegato di Salerno, I-84084 Fisciano (SA), Italy.  \label{addr2}}
\date{Received: date / Accepted: date}

\maketitle

\begin{abstract}
The (inverse) $\beta$-decay of uniformly accelerated
protons ($p\rightarrow n\hspace{0.2mm}+\hspace{0.2mm}
e^{+}\hspace{0.2mm}+\hspace{0.2mm}\nu_e$) has been 
recently analyzed in the context of two-flavor neutrino 
mixing and oscillations. It has been shown that the decay 
rates as measured by an inertial and comoving observer 
are in agreement, provided that: $i)$ the \emph{thermal nature}
of the accelerated vacuum (Unruh effect) is taken into account; $ii)$
the asymptotic behavior of neutrinos is described
through \emph{flavor} (rather than mass) \emph{eigenstates}; 
$iii)$ the Unruh radiation is made up of 
\emph{oscillating neutrinos}. Here we extend the above 
considerations to a more realistic scenario including 
three generations of Dirac neutrinos. By following the 
outlined recipe, we find that the equality between the two 
rates still holds true, confirming that mixing is perfectly 
consistent with the General Covariance of Quantum Field 
Theory. Notably, we prove that the analysis of CP violation 
in neutrino oscillations provides a further solid argument 
for flavor states as fundamental representation of asymptotic neutrino states.
Our approach is finally discussed in
comparison with the other treatments  
appeared in literature.
\end{abstract}

 \vskip -1.0 truecm
\maketitle

\section{Introduction}
 It is well-known that physical laws for accelerated systems
are far more subtle than the corresponding equations for  
inertially moving objects. 
In Classical Mechanics, for instance, one can
hold onto Newton's law $\textbf{F}=m\hspace{0.2mm}\textbf{a}$
for an observer trapped in a free-falling elevator or
rotating on a merry-go-round, provided
that some extra ``fictitious'' forces are introduced. 
In the same way, a charged particle undergoing 
an acceleration does radiate photons (whereas an inertial 
particle does not), 
the rate of which is predicted by Classical Electrodynamics 
to be proportional to the square of the acceleration~\cite{Jack}. 
In this context, the question naturally arises as to whether 
similar inertial effects also
come into play in a purely quantum realm.

Along this line, in 1976 Unruh found out that a uniformly 
accelerated (Rindler) observer experiences in the inertial (Minkowski)
vacuum a thermal bath of particles at temperature~\cite{Unruh}
\be
\label{UT}
T_{\mathrm U}=\frac{a}{2\pi}\,, 
\ee
where $a$ is the magnitude of the proper acceleration. 
This confirmed previous results about
the observer-dependence of the particle quantum concept even in
the absence of gravity~\cite{Full,Davies:1974th},
providing a flat-counterpart of the best-known Hawking effect~\cite{Hawk}.

Notwithstanding the large number of
theoretical applications
and the experimental efforts made so far~\cite{Crispi}, direct
evidences of Unruh radiation
are still lacking, thereby opening up a lively debate on its actual existence~\cite{doubt}, 
even through the study of analogue models~\cite{Liberati,Iorio,CapUnr}. 
Against the skepticism, however, a virtual confirmation of Unruh effect  
was elegantly proposed in the context of the  
inverse $\beta$-decay in Ref.~\cite{Matsas}, where it was 
shown that inertial and co-accelerated observers would draw
incompatible conclusions about the stability of non-inertial protons 
if the vacuum radiation were not taken into account.
In light of this, there is no question that the Unruh effect turns out to be
mandatory for Quantum Field Theory (QFT) 
as well as fictitious forces are for Classical Mechanics, since
both of them are required to preserve the
internal consistency of successfully tested 
theories when investigated in accelerated frames.

The intimate connection between the Unruh effect and 
the inverse $\beta$-decay was first addressed in a toy model 
in Ref.~\cite{Ginz,Mull}, assuming all involved particles to be scalars, 
and then analyzed within a more rigorous framework 
with Dirac fields in Refs.~\cite{SUZU,SUZU2}. Surprisingly, only 
recently it was studied in connection with neutrino flavor 
mixing and oscillations~\cite{Ahluw,NMIBD,NOUR,Cozzella}, with conflicting results on the very nature of asymptotic neutrino states being reached. 
In these works, a preliminary description including only two flavors was considered.

Starting from the outlined scenario, in what follows
we discuss the Unruh effect and revisit the inverse $\beta$-decay
with mixed neutrinos in a more realistic three-flavor setting. 
By explicit calculation, 
we show that a covariant treatment consistent
with the phenomena of mixing and oscillations
unavoidably implies the choice of 
flavor (rather than mass) eigenstates for asymptotic neutrinos, 
as well as the occurrence of flavor oscillations even in the Unruh thermal bath.
The obtained result is corroborated by very straightforward 
considerations on the necessity to allow for CP asymmetry in 
processes involving neutrino oscillations -- a feature which
mass eigenstates would fail to pinpoint. Based on these arguments, 
we also speculate on the possibility to have
a non-trivial asymmetry between the Unruh baths 
experienced by the accelerated proton and antiproton, respectively. 

The remainder of the work is organized as follows:
in Sec.~\ref{GC} we set the stage
for the study of the inverse $\beta$-decay.
Sec.~\ref{LF} is devoted
to the evaluation of the decay rate in the laboratory frame.
The same calculation
is independently performed from the point of view
of a comoving observer in Sec.~\ref{CF}. 
We show that the two results are
in full agreement, contrary to previous claims
of Ref.~\cite{Ahluw}. Sec.~\ref{Inconst} concerns  
a discussion on the incompatibility 
between the mass asymptotic representation and CP-violation
effects. Closing remarks are contained in Sec.~\ref{DeC}.
Throughout the paper, we shall use the Minkowski metric with the timelike signature and natural units $k_{\mathrm{B}}=\hslash=c=1$.

\section{Inverse $\beta$-decay and neutrino mixing: general considerations}
\label{GC}
Despite the common belief, the lifetime of a particle
cannot be considered among its inherent and characteristic properties.
The most eloquent example is provided by the proton, 
which is a \emph{stable} bound state of quarks, at least    
according to the predictions of the Standard Model. 
In Ref.~\cite{Ginz,Mull}, indeed, it was 
argued that the lifetime $\tau_p$ of the proton may significantly decrease if
we expose it to a large acceleration $a$. This was
rigorously shown in Refs.~\cite{SUZU,SUZU2}, where the inverse
$\beta$-decay 
\be
\label{lab}
(\mathrm{i})\quad p\myeq n\hspace{0.2mm}+\hspace{0.2mm}
e^{+}\hspace{0.2mm}+\hspace{0.2mm}\nu_e 
\ee 
was analyzed in both
the laboratory and comoving frames, obtaining 
non-vanishing (equal) results for the proton decay rate
$\Gamma\sim \tau_p^{-1}$.

The study of the interaction~\eqref{lab} proceeds in 
a straightforward way if we regard 
the proton $|p\rangle$ and neutron $|n\rangle$ 
as unexcited and excited states of a two-level system, 
the nucleon, whose Hamiltonian obeys the relations 
\begin{equation}
\hat H|p\rangle=m_p|p\rangle\,,\quad \hat H|n\rangle=m_n|n\rangle\,,
\end{equation}
where $m_{p(n)}$ is the rest mass of the proton (neutron).
Furthermore, we require that the momenta of 
both the positron $|e^+\rangle$ and neutrino $|\nu_e\rangle$ 
satisfy the condition $|\textbf{k}_{e^+(\nu_e)}|\ll m_p,m_n$, 
so that the fermion emission does not
change the four-velocity of the hadrons appreciably (\emph{no-recoil
approximation}). Within this semiclassical framework, it is reasonable to 
suppose that the nucleon system
will move along a well-defined
trajectory, the Rindler hyperbola, which indeed
describes a uniformly accelerated motion. By assuming the acceleration
to be directed along the $z$-axis, the associated
current can be written as
\be
\label{j1}
\hat{J}_{h,\la}=\hat{q}(\tau)u_{\la}\delta(x)\delta(y)\delta(u-1/a)\,,
\ee
where $\hat{q}(\tau)=e^{i\hat H\tau}\hat q(0)e^{-i\hat H\tau}$ is the 
monopole operator and $G_F=|\langle n|\hat q(0)|p\rangle|$ is the Fermi constant.
In the above expression, we have denoted by  $v$ and $\tau=v/a$ the Rindler 
time coordinate and proper time of the nucleon, respectively. The Dirac delta fixes
the spatial coordinate $u$ to the value $1/a$, which 
identifies the
Rindler trajectory\footnote{Note that the 
Rindler coordinates $(v,x',y',u)$ are related 
to the corresponding 
Minkowski coordinates $(t,x,y,z)$ by $t=u\sinh{v}$, $x'=x$, 
$y'=y$, $z=u\cosh{v}$.}. The nucleon four-velocity reads
$u^{\la}=(a, 0, 0, 0)$ and $u^{\la}=(\sqrt{a^2t^2+1}, 0, 0, a\hspace{0.2mm}t)$ 
in Rindler and Minkowski coordinates. 
 
In turn, leptons are treated as Dirac quantum fields 
with a current given by
\be
\label{j2}
\hat{J}^{\la}_l=\sum_{\ell=e,\mu,\tau}\left(\hat{\overline{\Psi}}_{\nu_\ell}\gamma^{\la}\hat{\Psi}_{\ell} + \hat{\overline{\Psi}}_{\ell}\gamma^{\la}\hat{\Psi}_{\nu_\ell}\right),
\ee 
where $\hat{\Psi}_{\ell}$ ($\hat\Psi_{\nu_\ell})$
is the charged lepton (neutrino) Dirac field and $e,\mu,\tau$
label the three lepton flavors. Rigorously speaking, 
we should consider a current including also the axial term.
As shown in Ref.~\cite{Matsas}, however, 
the above oversimplification does not affect the 
overall validity of our analysis.

By resorting to Eqs.~\eqref{j1} and~\eqref{j2},
the Fermi-like effective action describing the interaction
takes the form
\be
\label{eqn:Fermiaction}
 \hat{S}_{I}\,=\,\int d^{4}x\hspace{0.3mm}\sqrt{-g}\hspace{0.3mm}\hat{J}_{h,\la}\hspace{0.2mm}\hat{J}^{\la}_l,
 \ee
where $g=\mathrm{det}(g_{\mu\nu})$ and $\gamma^\la$ 
are the gamma matrices in the
Dirac representation (see Ref.~\cite{Itzykson}). 

In the simplest extended version of the Standard Model, it is
well-known that neutrinos weakly interact
with charged leptons in flavor
eigenstates $|\nu_\ell\rangle$~\cite{BilPont}, which 
are superpositions of mass states  
$|\nu_j\rangle$ $(j=1,2,3)$ determined by the transformation 
\begin{equation}
\label{eqn:U}
\begin{pmatrix}
\vspace{1mm}
|\nu_e\rangle  \\
\vspace{1mm}
|\nu_\mu\rangle\\
|\nu_\tau\rangle
\end{pmatrix}\,=\,U\hspace{-0.5mm}\begin{pmatrix}
\vspace{1mm}
|\nu_1\rangle  \\
\vspace{1mm}
|\nu_2\rangle\\
|\nu_3\rangle
\end{pmatrix},
\end{equation} 
where $U$ is the 
Pontecorvo-Maki-Nakagawa-Sakata (PMNS) \cite{PMNS} matrix 
in the standard parameterization~\cite{Beringer}
\begin{equation}
\label{PMM}
\small
U=
\begin{pmatrix}
\vspace{1mm}
c_{12}\hspace{0.4mm}c_{13}\,\,
&s_{12}\hspace{0.4mm}c_{13}\,\,
&s_{13}\hspace{0.3mm}e^{-i\delta}\\
\vspace{1mm}
-s_{12}\hspace{0.4mm}c_{23}-c_{12}\hspace{0.4mm}s_{23}\hspace{0.4mm}s_{13}\hspace{0.3mm}e^{i\delta}\,\,
&c_{12}\hspace{0.4mm}c_{23}-s_{12}\hspace{0.4mm}s_{23}\hspace{0.4mm}s_{13}\hspace{0.3mm}e^{i\delta}\,\,
&s_{23}\hspace{0.4mm}c_{13}\\
s_{12}\hspace{0.4mm}s_{23}-c_{12}\hspace{0.4mm}c_{23}\hspace{0.4mm}s_{13}\hspace{0.3mm}e^{i\delta}\,\,
&-c_{12}\hspace{0.4mm}s_{23}-s_{12}\hspace{0.4mm}c_{23}\hspace{0.4mm}s_{13}\hspace{0.3mm}e^{i\delta}\,\,
&c_{23}\hspace{0.4mm}c_{13}
\end{pmatrix}.
\end{equation}
\normalsize
Here $c_{jk}=\cos\theta_{jk}$, $s_{jk}=\sin\theta_{jk}$, 
$\theta_{jk}$ is the $\nu_j$-$\nu_k$ mixing angle 
and $\delta$ is the CP-violating phase.

By using the $S$-matrix framework, 
in the next Section we evaluate the transition probability
for the process~\eqref{lab} by assuming that the
states of neutrino far before and after the interaction 
are those with definite flavor~\cite{BV95, ChengYang, Ji, Castiglioncello2018}.
 Calculations
are performed at tree level both in the laboratory and comoving frames, 
with emphasis on the mandatory r\^ole of the Unruh effect 
for the consistency of the two approaches.

\section{Inverse $\beta$-decay in the laboratory frame}
\label{LF}
\begin{figure}[t]
\centering
{\resizebox{6.1cm}{!}{\includegraphics{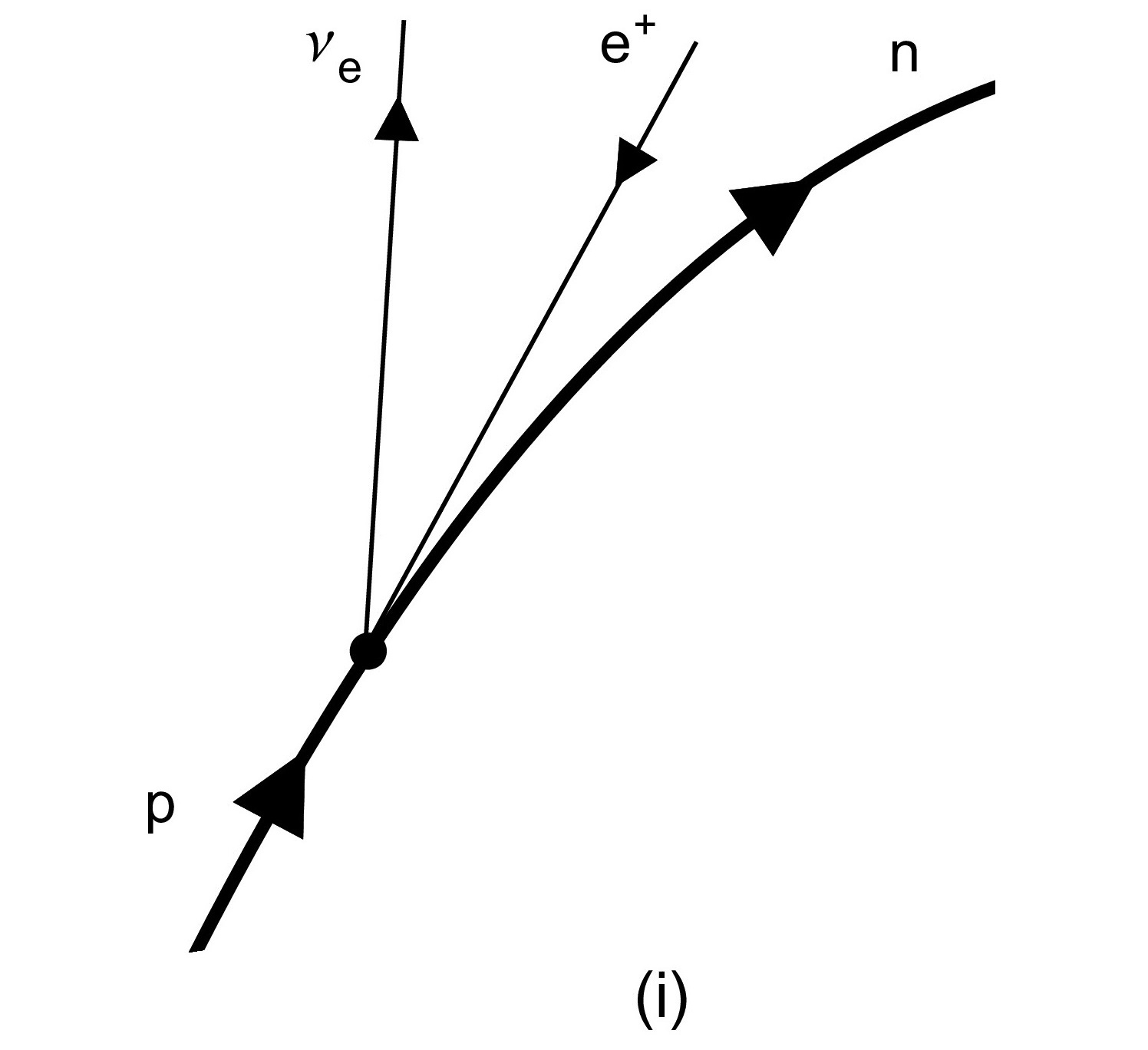}}}
\caption{Inverse $\beta$-decay in the laboratory frame in the absence of flavor oscillations (time flows in the vertical direction).}
\label{in}
\end{figure}

The inverse $\beta$-decay as seen by a Minkowski (inertial)
observer is given by Eq.~\eqref{lab} (see also Fig.~\ref{in}).
In order to evaluate the transition rate, we quantize the 
fermion fields in the standard way~\cite{NMIBD}
\be
\label{inertexp}
\hat{\Psi}(t,\textbf{x})\ =\ \sum_{\sigma=\pm}\int d^3k\left[\hat{b}_{\textbf{k}\sigma}\psi_{\bf{k}\sigma}^{(+\omega)}\,+\,\hat{d}_{\bf{k}\sigma}^{\dagger}\psi_{-\textbf{k}-\sigma}^{(-\omega)}\right],
\ee
where $\textbf{x}\equiv(x,y,z)$ (for simplicity, 
we have omitted the spacetime dependence of the spinors in the r.h.s. of the expansion). 
We have denoted by $\hat{b}_{\textbf{k}\sigma}$ ($\hat{d}_{\textbf{k}\sigma}$) the canonical annihilation operators of particles (antiparticles) with momentum $\textbf{k}\equiv(k^x, k^y, k^z)$, polarization $\sigma=\pm$, frequency $\omega=\sqrt{\textbf{k}^{2}+m^{2}}$ and mass $m$.  The modes $\psi_{\textbf{k}\sigma}^{(\pm\omega)}$ are positive and negative energy solutions of the Dirac equation in Minkowski spacetime. They are given by
\be
\label{modes}
\psi_{\textbf{k}\sigma}^{(\pm\omega)}(t, \textbf{x}) \ =\ \frac{e^{i(\mp\omega t\,+\,\textbf{k}\cdot\textbf{x})}}{2^2\hspace{0.2mm}\pi^{\frac{3}{2}}}\hspace{0.2mm}\hspace{0.2mm}u_{\sigma}^{(\pm\omega)}(\textbf{k})\,,
\ee 
where
\begin{eqnarray}
u_{+}^{(\pm\omega)}(\textbf{k})&=& \frac{1}{\sqrt{\omega(\omega\pm m)}}
\begin{pmatrix}
m\,\pm\,\omega \\
\vspace{1mm}
0 \\
\vspace{1mm}
k^z \\
\vspace{1mm}
k^x\,+\,ik^y
\end{pmatrix},\\[2.5mm]
u_{-}^{(\pm\omega)}(\textbf{k})&=& \frac{1}{\sqrt{\omega(\omega\pm m)}}
\begin{pmatrix}
0 \\
\vspace{1mm}
m\,\pm\,\omega \\
\vspace{1mm}
k^x\,-\,ik^y \\
\vspace{1mm}
-\hspace{0.2mm}k^z
\end{pmatrix}.
\end{eqnarray}
With the above definition, one can easily prove
that the modes $\psi_{\textbf{k}\sigma}^{(\pm\omega)}$ are orthonormal with respect 
to the inner product~\cite{Birrell} 
\begin{eqnarray}
\label{innprod}
\non
\left\langle\psi_{\textbf{k}\sigma}^{(\pm\omega)},\psi_{\textbf{k}'\sigma'}^{(\pm\omega')}\right\rangle&\equiv&\int_{\Sigma}d\Sigma_{\lambda}\,\overline{\psi}_{\textbf{k}\sigma}^{(\pm\omega)}\gamma^{\lambda}\psi_{\textbf{k}'\sigma'}^{(\pm\omega')}\\[2mm]
&=&\delta_{\sigma\sigma'}\delta^3(\textbf{k}-\textbf{k}')\delta_{\pm\omega\pm\omega'},
\end{eqnarray}
where $\overline{\psi}\,=\,\psi^{\dagger}\gamma^{0}$, $d\Sigma_{\lambda}\,=\,n_{\lambda}d\Sigma$ and $n_{\lambda}$ is a unit vector orthogonal to the hypersurface $\Sigma$ of constant $t$. 

In the $S$-matrix formalism, the transition amplitude for the process~\eqref{lab} reads\footnote{The consistency of the $S$-matrix formalism with flavor asymptotic states has been questioned several times in literature~\cite{Giunti:2003dg,Giuntibook,Cozzella}. In spite of this, one can prove that such an approach 
is well-posed both physically (since its predictions are
in agreement with the ones of the Standard Model) and mathematically (as the asymptotic $t\rightarrow\pm\infty$ limits
do not entail any technical problem in the calculation of transition amplitudes)~\cite{ChengYang}.}~\cite{NMIBD}
\begin{eqnarray}\label{tramp}
\non
\mathcal{A}^{\mathrm{(i)}}&\equiv&\langle n|\otimes\langle e^{+},\nu_{e}|\hat{S}_{I}|0\rangle\otimes|p\rangle\\[2mm]
&=&\frac{G_F}{2^4\pi^3}\sum_{j=1}^3{\lf|U_{ej}\ri|}^2\, \mathcal{I}_{\sigma_\nu\sigma_e}(\omega_{\nu_j},\omega_e)\,,
\end{eqnarray}
where the latin number in the superscript of the l.h.s. labels
the process under consideration and $U_{\ell j}$ ($\ell=e,\mu,\tau$) is the generic element of the PMNS matrix~\eqref{PMM}. The function 
$\mathcal{I}_{\sigma_\nu\sigma_e}$ is defined as
\begin{eqnarray}
\label{I}
\non
\mathcal{I}_{\sigma_\nu\sigma_e}(\omega_{\nu_j}, \omega_e)&=&\int_{-\infty}^{+\infty}d\tau\, u_{\lambda}\left[\bar{u}_{\sigma_\nu}^{(+\omega_{\nu_j})}\gamma^{\lambda}{u}_{-\sigma_e}^{(-\omega_e)}\right]\\[2mm]
&&\hspace{-23mm}\times\, e^{i\big[\Delta m\hspace{0.2mm}\tau\,+\,a^{-1}\left(\omega_{\nu_j}\,+\,\omega_{e}\right)\sinh a\tau\,-\,a^{-1}\left(k_\nu^z\,+\,k^z_{e}\right)\cosh a\tau\big]}\,. 
\end{eqnarray}
Note that, in the above calculation, 
we have implemented the PMNS transformation
on both neutrino state and field, the latter being transformed
as
\be
\label{field}
\psi_{\nu_e}(x)\,=\,\sum_{j=1}^3U^*_{ej}\hspace{0.3mm}\psi_{\nu_j}(x)\,,
\ee
and similarly for $\psi_{\nu_\mu}$ and $\psi_{\nu_\tau}$\footnote{
In Ref.~\cite{BV95} it has been shown 
that the structure of the field mixing
relation~\eqref{field} is that of a Bogoliubov transformation nested 
into the standard quantum mechanical rotation. As a result, the definition~\eqref{eqn:U} of mixed states turns out to be inconsistent with Eq.~\eqref{field}. 
Nevertheless, for relativistic neutrinos
and in the approximation of small mass
differences we are employing here, one can prove that  
Pontecorvo states well-approximate the
exact field theoretical states~\cite{Castiglioncello2018}, thus 
validating the simultaneous use of Eqs.~\eqref{eqn:U} and~\eqref{field}.}.
Moreover,  we have assumed equal
momenta and polarizations for neutrino states with definite
mass. 

Now, by plugging the amplitude~\eqref{tramp}
in the following expression of the scalar transition probability
per proper time $T$,
\be
\label{scaldeca}
\Gamma^{(\mathrm{i})}\,\equiv\,\frac{1}{T}\sum_{\sigma_e,\sigma_{\nu}}\int d^3k_\nu\int d^3k_e\,
\big|\mathcal{A}^{{(\mathrm{i})}}\big|^2\,,
\ee
we get 
\begin{eqnarray}
\label{tpppt}
\non
\Gamma^{(\mathrm{i})}&=&{|U_{e1}|}^4\,\,\Gamma_1\,+\,{|U_{e2}|}^4\,\,\Gamma_2\,+\,{|U_{e3}|}^4\,\,\Gamma_3\\[2mm]\non
&&+\,\Big({|U_{e1}|}^2\,{|U_{e2}|}^2\,\,\Gamma_{12}\,+\,{|U_{e1}|}^2\,{|U_{e3}|}^2\,\,\Gamma_{13}\\[2mm]
&&+\,{|U_{e2}|}^2\,{|U_{e3}|}^2\,\,\Gamma_{23}\,+\,\mathrm{c.c.}\Big).
\end{eqnarray}
Here, the contribution
\begin{eqnarray}
\label{integral}
\non
\Gamma_{j} &\equiv& \frac{1}{T}\,\frac{G_F^2}{2^8\pi^6}\sum_{\sigma_\nu,\sigma_e}\int d^3k_\nu\int d^3k_e\,{\big|\mathcal{I}_{\sigma_\nu\sigma_e}(\omega_{\nu_j},\omega_e)\big|}^2\\[2mm]\non
&=&\frac{G_F^2}{a\hspace{0.2mm}\pi^6e^{\pi\Delta m/a}}\int d^3 k_\nu\int d^3 k_e \left\{K_{2i\Delta m/a}\left(\frac{2(\omega_{\nu_j}+\omega_e)}{a}\right)\right.\\[2mm]
&&+\,\frac{m_{\nu_j} m_e}{\omega_{\nu_j}\omega_e}\,\mathrm{Re}\left[K_{2i\Delta m/a+2}\left(\frac{2(\omega_{\nu_j}+\omega_e)}{a}\right)\right]\bigg\}
\end{eqnarray}
represents the decay rate we would obtain by using  
$|\nu_j\rangle$ as asymptotic neutrino state, $K_{i\nu}(x)$ is the modified Bessel function of second kind, and the interference term
\begin{eqnarray}
\label{integraloffdiag}
\Gamma_{jk}&\equiv& \frac{1}{T}\,\frac{G_F^2}{2^8\pi^6}\\[2mm]\non
&&\times\hspace{-1mm}\sum_{\sigma_\nu,\sigma_e\hspace{-0.5mm}}\int d^3k_\nu\hspace{-0.5mm}\int d^3k_e\,\mathcal{I}_{\sigma_\nu\sigma_e}(\omega_{\nu_j},\omega_e)\,\mathcal{I}_{\sigma_\nu\sigma_e}^{\hspace{0.2mm}*}(\omega_{\nu_k},\omega_e)
\end{eqnarray}
arises from the coherent superpositions of
neutrino states with different masses. The explicit expression
of $\Gamma_{jk}$ is rather awkward to exhibit. We remand to Ref.~\cite{NMIBD}
for a more detailed treatment of this term. 

Some comments are in order here: first, we observe 
that the decay rate~\eqref{tpppt} 
is given by the
coherent sum of the $\Gamma_j$-components of
the inverse $\beta$-decay amplitudes. By contrast, there is 
no summation over $j$ in the corresponding result~(48) of Ref.~\cite{Cozzella},
where asymptotic neutrinos are assumed to be
mass eigenstates. 
Such a controversy is addressed more specifically in 
Refs.~\cite{ChengYang, Castiglioncello2018, NOUR}.
Furthermore, for $\theta_{jk}\rightarrow0$,  
we recover the result of Ref.~\cite{SUZU2},
where the inverse $\beta$-decay is analyzed in the absence of mixing.
Similar considerations
hold in the approximation of small mass differences, 
since both $\Gamma_{i}$ ($i=2,3$) and 
$\Gamma_{jk}$ reduce to $\Gamma_1$ and
\begin{eqnarray}
\label{prop}
{|U_{e1}|}^4\,+\,{|U_{e2}|}^4\,+\,{|U_{e3}|}^4\\[2mm]\non
&&\hspace{-33mm}+\,2\left({|U_{e1}|}^2\,{|U_{e2}|}^2\,+\,{|U_{e1}|}^2\,{|U_{e3}|}^2\,+\,{|U_{e2}|}^2\,{|U_{e3}|}^2\right)=1\,.
\end{eqnarray}
\begin{figure}[t]
{\resizebox{8.1cm}{!}{\includegraphics{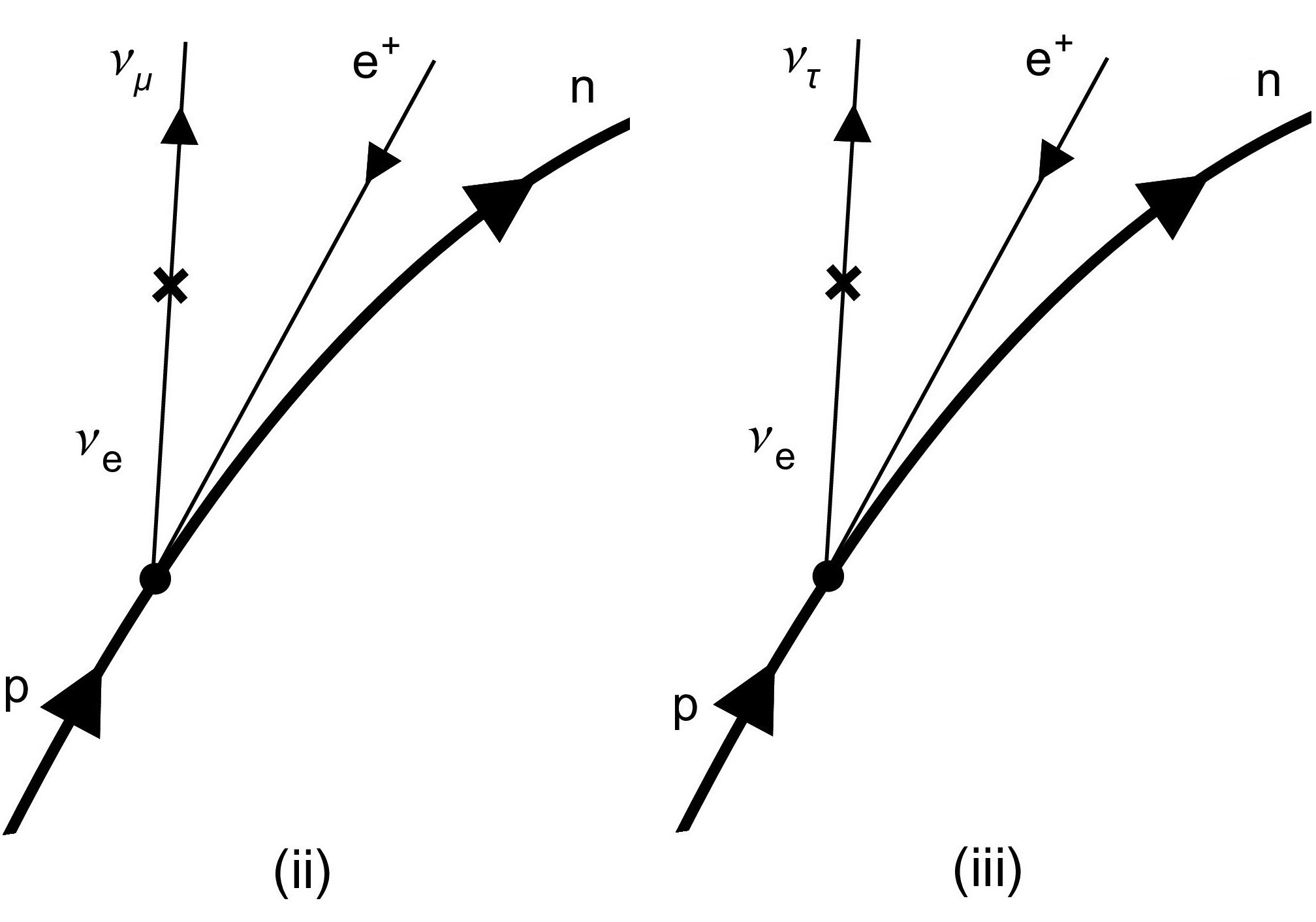}}}
\caption{Inverse $\beta$-decay in the laboratory frame in the presence of flavor oscillations (time flows in the vertical direction). The crosses denote the occurrence of the oscillation.}
\label{inosc}
\end{figure}
\begin{figure*}[t]
\centering
\hspace{0.8cm}
{\resizebox{13.5cm}{!}{\includegraphics{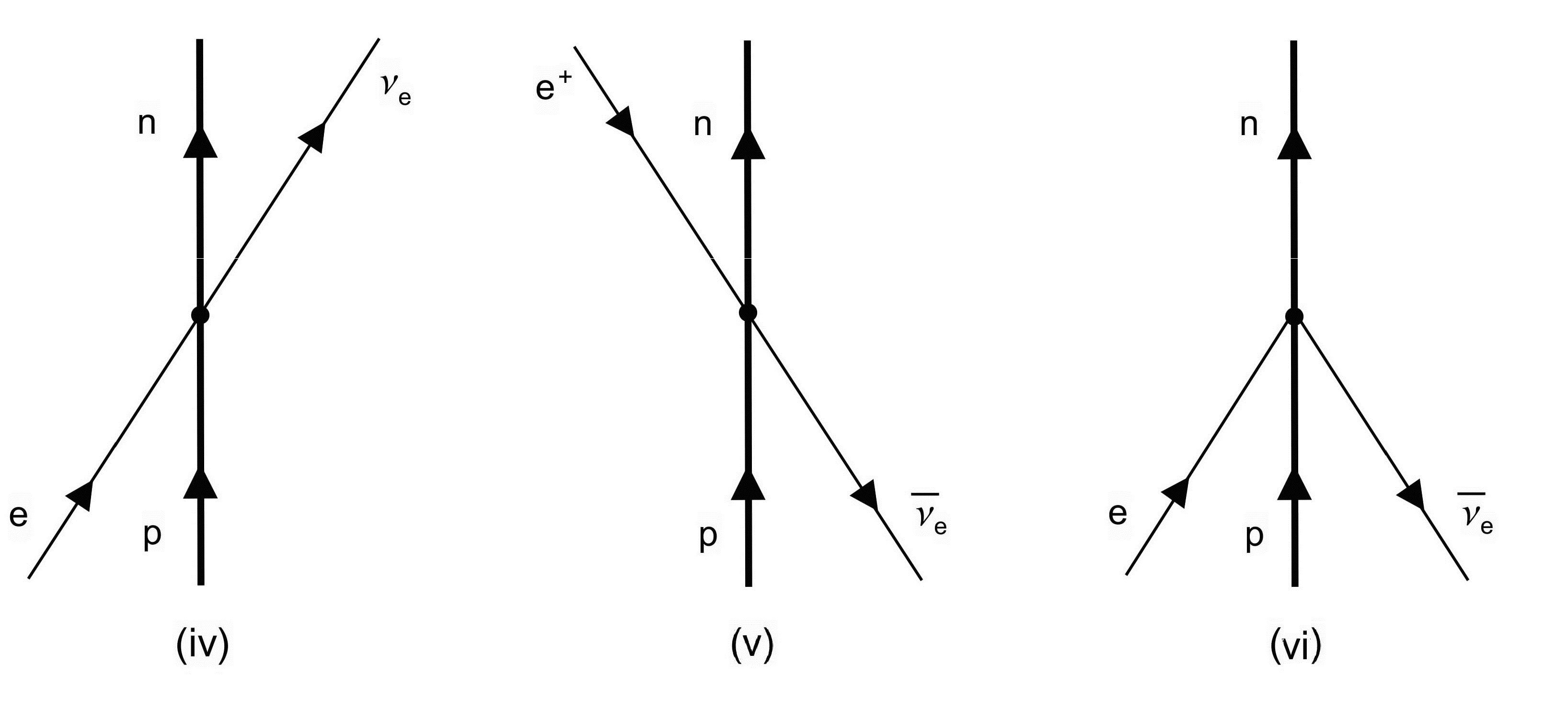}}}
\caption{Inverse $\beta$-decay in the comoving frame in the absence of flavor oscillations (time flows in the vertical direction). The three processes are allowed due to the absorption (emission) of particles from (to) the Unruh thermal bath.}
\label{accnoosc}
\end{figure*}
It is now worth noting that, due to the asymptotic effects of  
neutrino oscillations~\cite{NOUR}, 
the total rate for the inverse $\beta$-decay also gets 
non-vanishing contributions from the two
flavor-violating processes (see Fig.~\ref{inosc})
\begin{eqnarray}
\label{lab2}
(\mathrm{ii})\,\, &p\rightarrow n\hspace{0.2mm}+\hspace{0.2mm}
e^{+}\hspace{0.2mm}+\hspace{0.2mm}\nu_\mu\,,\\[2mm]
(\mathrm{iii})\,\, &p\rightarrow n\hspace{0.2mm}+\hspace{0.2mm}
e^{+}\hspace{0.2mm}+\hspace{0.2mm}\nu_\tau\,.
\label{lab3}
\end{eqnarray}
By using Eq.~\eqref{eqn:U}, one can show that 
the transition amplitudes for these channels
take the form
\begin{eqnarray}
\non
\mathcal{A}^{\mathrm{(ii)}}&\equiv&\langle n|\otimes\langle e^{+},\nu_{\mu}|\hat{S}_{I}|0\rangle\otimes|p\rangle\\[2mm]
&=&\frac{G_F}{2^4\pi^3}\sum_{j=1}^3U_{\mu j}^*\,U_{ej}\,\mathcal{I}_{\sigma_\nu\sigma_e}(\omega_{\nu_j},\omega_e)\,,
\end{eqnarray}
\begin{eqnarray}
\non
\mathcal{A}^{\mathrm{(iii)}}&\equiv&\langle n|\otimes\langle e^{+},\nu_{\tau}|\hat{S}_{I}|0\rangle\otimes|p\rangle\\[2mm]
&=&\frac{G_F}{2^4\pi^3}\sum_{j=1}^3U_{\tau j}^*\,U_{ej}\,\mathcal{I}_{\sigma_\nu\sigma_e}(\omega_{\nu_j},\omega_e)\,,
\end{eqnarray}
which lead to the following expressions for
the transition probabilities per proper time
\begin{eqnarray}
\label{mutau}
\non
\Gamma^{(\mathrm{ii})}&=&{|U_{\mu1}|}^2{|U_{e1}|}^2\,\,\Gamma_1\,+\,{|U_{\mu2}|}^2{|U_{e2}|}^2\,\,\Gamma_2\,+\,{|U_{\mu3}|}^2{|U_{e3}|}^2\,\,\Gamma_3\,\\[2mm]\non
&&+\Big(U_{\mu1}^*\,U_{e1}\,{U_{\mu2}\,U^*_{e2}}\,\,\Gamma_{12}\,+\,U_{\mu1}^*\,U_{e1}\,{U_{\mu3}\,U^*_{e3}}\,\,\Gamma_{13}\\[2mm]
&&+\,U_{\mu2}^*\,U_{e2}\,{U_{\mu3}\,U^*_{e3}}\,\,\Gamma_{23}\,+\,\mathrm{c.c.}\Big),
\end{eqnarray}
\begin{eqnarray}
\label{mutau2}
\non
\Gamma^{(\mathrm{iii})}&=&{|U_{\tau1}|}^2{|U_{e1}|}^2\,\,\Gamma_1\,+\,{|U_{\tau2}|}^2{|U_{e2}|}^2\,\,\Gamma_2\,+\,{|U_{\tau3}|}^2{|U_{e3}|}^2\,\,\Gamma_3\,\\[2mm]\non
&&+\Big(U_{\tau1}^*\,U_{e1}\,{U_{\tau2}\,U^*_{e2}}\,\,\Gamma_{12}\,+\,U_{\tau1}^*\,U_{e1}\,{U_{\tau3}\,U^*_{e3}}\,\,\Gamma_{13}\\[2mm]
&&+\,U_{\tau2}^*\,U_{e2}\,{U_{\tau3}\,U^*_{e3}}\,\,\Gamma_{23}\,+\,\mathrm{c.c.}\Big)\,.
\end{eqnarray}
Unlike $\Gamma^{(\mathrm{i})}$ in Eq.~\eqref{tpppt}, both $\Gamma^{(\mathrm{ii})}$ and $\Gamma^{(\mathrm{iii})}$
vanish for $\theta_{jk}\rightarrow0$ and/or for small mass differences, since
\begin{eqnarray}
\non
{|U_{\chi1}|}^2{|U_{e1}|}^2\,+\,{|U_{\chi2}|}^2{|U_{e2}|}^2\,+\,{|U_{\chi3}|}^2{|U_{e3}|}^2\\[2mm]\non
&&\hspace{-60mm}+\,\Big(U_{\chi1}^*\,U_{e1}\,{U_{\chi2}\,U^*_{e2}}\,+\,U_{\chi1}^*\,U_{e1}\,{U_{\chi3}\,U^*_{e3}}\\[2mm]
&&\hspace{-60mm}+\,\,U_{\chi2}^*\,U_{e2}\,{U_{\chi3}\,U^*_{e3}}\,+\,\mathrm{c.c.}\Big)\,=\,0\,,\quad\quad \chi=\mu,\tau.
\end{eqnarray}
This shows that $\Gamma^{(\mathrm{ii})}$ and $\Gamma^{(\mathrm{iii})}$ 
are pure interference
terms, whose origin is intimately related to the
non-trivial nature of neutrino mixing and oscillations.

Finally, by summing up the contributions in Eqs.~\eqref{tpppt},~\eqref{mutau} and~\eqref{mutau2}, the total decay rate becomes
\be
\label{totdecrat}
\Gamma^{\mathrm{lab}}\,\equiv\,\sum_{s=\mathrm{i}}^{\mathrm{iii}}\Gamma^{(s)}\,=\,
{|U_{e1}|}^2\,\,\Gamma_1\,+\,{|U_{e2}|}^2\,\,\Gamma_2\,+\,{|U_{e3}|}^2\,\,\Gamma_3\,.
\ee
From Eq.~\eqref{totdecrat} it arises 
that the sum over flavors of the inverse $\beta$-decay rates 
amounts to the weighted average 
over masses, with weights given by
the square modulus of the 
projections of $|\nu_e\rangle$ on $|\nu_j\rangle$ ($j=1,2,3$). The meaning
of this result can be illustrated as follows: let us consider the lepton 
charges for mixed neutrinos as derived from Noether's theorem. 
If we denote by $Q_j=\int d^3x\, \Psi_{\nu_j}^\dag(x)\Psi_{\nu_j}(x)$
the conserved charge for the neutrino field with mass $m_j$ 
and by $Q_\ell(t)=\int d^3x\,\Psi_{\nu_\ell}^\dag(x)\Psi_{\nu_\ell}(x)$
the (time-dependent) flavor charge for the field with definite flavor $\ell$, 
we simply have $Q=\sum_{j=1}^3 Q_j=\sum_{\ell=e,\mu,\tau} Q_\ell(t)$,
with $Q$ being the total charge of the system~\cite{Blasone:2001qa}. 
Beyond the pure mathematical equality,  
the physical interpretation of this relation is
non-trivial, as it states that the total lepton number is a conserved quantity
both in the presence and in the absence of flavor mixing.
On the left side (i.e. when mixing is not taken into account), such quantity 
is given by the sum of three separately conserved family lepton numbers; conversely, 
on the right side (i.e. when mixing is included) it is 
obtained by summing up three non-conserved flavor charges, which are indeed associated to the phenomenon of neutrino oscillations.

\section{Inverse $\beta$-decay in the comoving frame}
\label{CF}
From the viewpoint of an observer comoving with the proton, 
the process~\eqref{lab} is clearly forbidden by energy conservation. 
According to such an observer, however, the Minkowski vacuum appears
as a thermal bath of virtual particles with which the proton can interact. 
Consequently, the following new channels become accessible (see Fig.~\ref{accnoosc}):
\begin{subequations}
\label{comov}
\begin{align}
\label{31a}
&\mathrm{(iv)}\quad p\,+\,e^{-}\,\rightarrow\, n\,+\,\nu_e\,,\\[2mm]
\label{31b}
&\mathrm{(v)}\quad p\,+\,\overline{\nu}_e\,\rightarrow\, n\,+\,e^{+}\,,\\[2mm]
\label{31c}
&\mathrm{(vi)}\quad p\,+\,e^{-}\,+\,\overline{\nu}_e\,\rightarrow\, n\,,
\end{align}
\end{subequations}
i.e. the proton at rest is allowed to decay
due to the absorption of an electron (Eq.~\eqref{31a}), an antineutrino (Eq.~\eqref{31b}) and
both an electron and an antineutrino (Eq.~\eqref{31c}) from the thermal bath.

In order to compute the total transition probability, 
let us remind that the proper way to quantize fields 
for a uniformly accelerated observer  
is the Rindler-Fulling
scheme, according to which~\cite{NMIBD} 
\be 
\label{Rindexp}
\hat{\Psi}(v,\textbf{x})=\sum_{\sigma=\pm}\int_{0}^{+\infty}\hspace{-2mm}d\omega\int d^2k\left[\hat{b}_{\textbf{w}\sigma}\psi^{(+\omega)}_{\textbf{w}\sigma}\,+\,\hat{d}_{\textbf{w}\sigma}^{\dagger}\psi^{(-\omega)}_{\textbf{w}-\sigma}\right],
\ee
where $\textbf{x}\equiv(x,y,u)$, $\textbf{w}\equiv(\omega, k^x, k^y)$.
Here, we have denoted by
$\hat{b}_{\textbf{w}\sigma}$ $(\hat{d}_{\textbf{w}\sigma})$ 
the canonical annihilation operators of Rindler particles (antiparticles) with transverse momentum $k\equiv(k^x, k^y)$, polarization $\sigma=\pm$ and frequency $\omega>0$. Note that, contrary to the Minkowski case, this frequency is independent of  the mass of Rindler quanta, since it does not satisfy any dispersion relation.  The positive/negative energy solutions of the Dirac equation in Rindler spacetime take the form
\be
\label{modesRind}
\psi_{\textbf{w}\sigma}^{(\omega)}(v, \textbf{x})\ =\ \frac{e^{i(-\omega v/a\,+\,k_\alpha x^\alpha)}}{{(2\pi)}^{\frac{3}{2}}}\hspace{0.2mm}u_{\sigma}^{(\omega)}\hspace{0.2mm}(u,\textbf{w})\,,\quad \alpha=1,2\,,
\ee 
where
\begin{eqnarray}
\label{Rindmodes}
u_{+}^{(\omega)}(u,\textbf{w})&\hspace{-0.3mm}=\hspace{-0.3mm}&
N\hspace{-0.4mm}\begin{pmatrix}
i\hspace{0.2mm}l\hspace{0.1mm}K_{i\omega/a-1/2}(u\hspace{0.2mm}l)\,+\,m\hspace{0.2mm}K_{i\omega/a+1/2}(u\hspace{0.2mm}l) \\[2.5mm]
-(k^x+ik^y)K_{i\omega/a+1/2}(u\hspace{0.2mm}l) \\[2.5mm]
i\hspace{0.2mm}l\hspace{0.1mm}K_{i\omega/a-1/2}(u\hspace{0.2mm}l)\,-\,m\hspace{0.2mm}K_{i\omega/a\,+\,1/2}(u\hspace{0.2mm}l) \\[2.5mm]
-(k^x+ik^y)K_{i\omega/a+1/2}(u\hspace{0.2mm}l)
\end{pmatrix}\hspace{-0.8mm},\\[5mm]
u_{-}^{(\omega)}(u,\textbf{w})&\hspace{-0.3mm}=\hspace{-0.3mm}&
N\hspace{-0.4mm}\begin{pmatrix}
(k^x-ik^y) K_{i\omega+1/2}(u\hspace{0.2mm}l) \\[2.5mm]
i\hspace{0.3mm}l\hspace{0.1mm}K_{i\omega/a-1/2}(u\hspace{0.2mm}l)\,+\,m\hspace{0.2mm}K_{i\omega/a+1/2}(u\hspace{0.2mm}l) \\[2.5mm]
-(k^x-ik^y) K_{i\omega+1/2}(u\hspace{0.2mm}l) \\[2.5mm]
-i\hspace{0.2mm}l\hspace{0.1mm}K_{i\omega/a-1/2}(u\hspace{0.2mm}l)\,+\,m\hspace{0.2mm}K_{i\omega/a+1/2}(u\hspace{0.2mm}l) 
\end{pmatrix}
\end{eqnarray}
with $N\equiv\sqrt{\frac{a\cosh(\pi\omega/a)}{\pi l}}$ and $l\equiv\sqrt{(k^x)^2+(k^y)^2+m^2}$. 

Let us now sketch the procedure
to evaluate the decay rate for the process 
$(\mathrm{iv})$ in Eq.~\eqref{31a}; similar considerations can be straightforwardly generalized to the channels $(\mathrm{v})$ and $(\mathrm{vi})$.
First, by using the field expansion~\eqref{Rindexp} and rotating the
neutrino state and field according to 
Eq.~\eqref{eqn:U} and~\eqref{field}, the transition 
amplitude can be expressed as~\cite{NMIBD}
\begin{eqnarray}
\label{first}
\non
\mathcal{A}^{(\mathrm{iv})}&\equiv&\left\langle n\right|\otimes\langle\nu_{e}|\hspace{0.2mm}\hat{S}_{I}\hspace{0.2mm}|e^{-}\rangle\otimes\left|p\right\rangle\\[2.5mm]
&=&\frac{G_F}{(2\pi)^2}\sum_{j=1}^3{\lf|U_{ej}\ri|}^2\,\mathcal{J}^{(j)}_{\sigma_\nu\sigma_e}(\omega_\nu, \omega_e)
\end{eqnarray}
where 
\be
\label{J}
\mathcal{J}^{(j)}_{\sigma_\nu\sigma_e}(\omega_\nu, \omega_e)\,=\,\delta\big(\omega_e-\omega_{\nu}-\Delta m\big)\,\bar{u}_{\sigma_{\nu}}^{(\omega_{\nu})}\gamma^0 u_{\sigma_{e}}^{(\omega_{e})}\,,
\ee
and we have assumed equal frequencies, transverse momenta and polarizations
for neutrino states with definite
mass. As a next step, 
it should be considered that, due to the Unruh
effect, the probability that the proton absorbs 
a lepton of frequency $\omega$ from the thermal bath
is given by the Fermi-Dirac distribution 
\be
n_F(\omega)\,=\,{({e^{\omega/T_{\mathrm{U}}}+1})}^{-1}\,,
\ee 
(similarly, the probability to emit a particle to the bath reads $\widetilde{n}_F(\omega)=1-n_{F}(\omega)$), where the temperature $T_{\mathrm{U}}$
is defined as in Eq.~\eqref{UT}. 
Calculations are finalized by multiplying the above
thermal factors by the squared modulus of the amplitude~\eqref{first}, then 
integrating over the Rindler
momentum-space volume $dV_{k,R}=d\omega_\nu d\omega_e d^2k_\nu d^2k_e$ and summing over the leptons' polarizations $\sigma_{\nu}, \sigma_e$ (see Ref.~\cite{NMIBD} for explicit calculations).

If we now follow the above recipe for all three processes 
in Eqs.~\eqref{comov} and add up the resulting expressions for the
decay rates, we obtain  
\begin{eqnarray}
\non
\label{TIR}
\Gamma^{(\mathrm{iv})}\,+\,\Gamma^{(\mathrm{v})}\,+\,\Gamma^{(\mathrm{vi})}&=&{|U_{e1}|}^4\,\,\widetilde\Gamma_1\,+\,{|U_{e2}|}^4\,\,\widetilde\Gamma_2\,+\,{|U_{e3}|}^4\,\,\widetilde\Gamma_3\\[2.5mm]\non
&&\hspace{-30mm}+\left({|U_{e1}|}^2\,{|U_{e2}|}^2\,\,\widetilde\Gamma_{12}\,+\,{|U_{e1}|}^2\,{|U_{e3}|}^2\,\,\widetilde\Gamma_{13}\right.\\[2mm]
&&\hspace{-30mm\,+\,{|U_{e2}|}^2\,{|U_{e3}|}^2\,\,\widetilde\Gamma_{23}}\,+\,\mathrm{c.c.}\Big),
\end{eqnarray}
where 
\begin{eqnarray}
\label{integralbis}
\non
\widetilde\Gamma_{j}&\equiv& \frac{2\,G_{F}^{2}}{a^2\pi^7e^{\pi\Delta m/a}}\hspace{-0.5mm}\int \hspace{-0.7mm}d\omega\hspace{0.4mm}\Bigg\{\hspace{-1mm}\int d^2k_{\nu}\,l_{\nu_j}\Bigl|K_{i(\omega-\Delta m)/a+1/2}\left(\frac{l_{\nu_j}}{a}\right)\Bigr|^2\\[2mm]\non
&&\times\int d^2k_{e}\,l_{e}\Bigl|K_{i\omega/a+1/2}\left(\frac{l_{e}}{a}\right)\Bigr|^2\\[2mm]\non
&&+\,\,m_{\nu_j}m_e\,\mathrm{Re}\hspace{0.2mm}\left[\int d^2k_{\nu}\,K^2_{i(\omega-\Delta m)/a-1/2}\left(\frac{l_{\nu_j}}{a}\right)\right.\\[2mm]
&&\left.\,\times\int d^2k_eK^2_{i\omega/a+1/2}\left(\frac{l_{e}}{a}\right)\right]\Bigg\}, 
\end{eqnarray}
\vspace{1mm}
\begin{eqnarray}\nonumber
\label{int1212}
\widetilde\Gamma_{jk}&=&\frac{G_{F}^{2}}{a^2\pi^7e^{\pi\Delta m/a}}\int \hspace{-0.5mm}d\omega\,\Biggl\{\,\int d^2k_e\,l_e\Bigl|K_{i\omega/a+1/2}\left(\frac{l_e}{a}\right)\Bigr|^2\\[1.5mm]\non
&&\times\int \frac{d^2k_\nu}{\sqrt{l_{\nu_j}l_{\nu_k}}}\,\Big({(k_\nu^x)}^2+{(k_\nu^y)}^2\,+\,m_{\nu_j}m_{\nu_k}\,+\,l_{\nu_j}l_{\nu_k}\Big)\\[1.5mm]\nonumber
&&\times\,\left[K_{i(\omega-\Delta m)/a+1/2}\left(\frac{l_{\nu_j}}{a}\right)K_{i(\omega-\Delta m)/a-1/2}\left(\frac{l_{\nu_k}}{a}\right)\right]\\[2mm]\non
&&+\,\, \frac{m_e}{\sqrt{l_{\nu_j}l_{\nu_k}}}\int d^2k_e\int d^2k_\nu\big(l_{\nu_j}m_{\nu_k}\,+\,l_{\nu_k}m_{\nu_j}\big)\\[1.5mm]\non
&&\times\,\left[K^2_{i\omega/a+1/2}\left(\frac{l_e}{a}\right)K_{i(\omega-\Delta m)/a-1/2}\left(\frac{l_{\nu_j}}{a}\right)\right.\\[1.5mm]
&&\left.\times\, K_{i(\omega-\Delta m)/a-1/2}\left(\frac{l_{\nu_k}}{a}\right)\right]\Biggr\}\,.
\end{eqnarray}

Again, for vanishing mixing angles and/or small mass differences, 
we recover the result of Ref.~\cite{SUZU2} (see the discussion
before Eq.~\eqref{prop}).

\begin{figure*}[t]
\centering
{\resizebox{12cm}{!}{\includegraphics{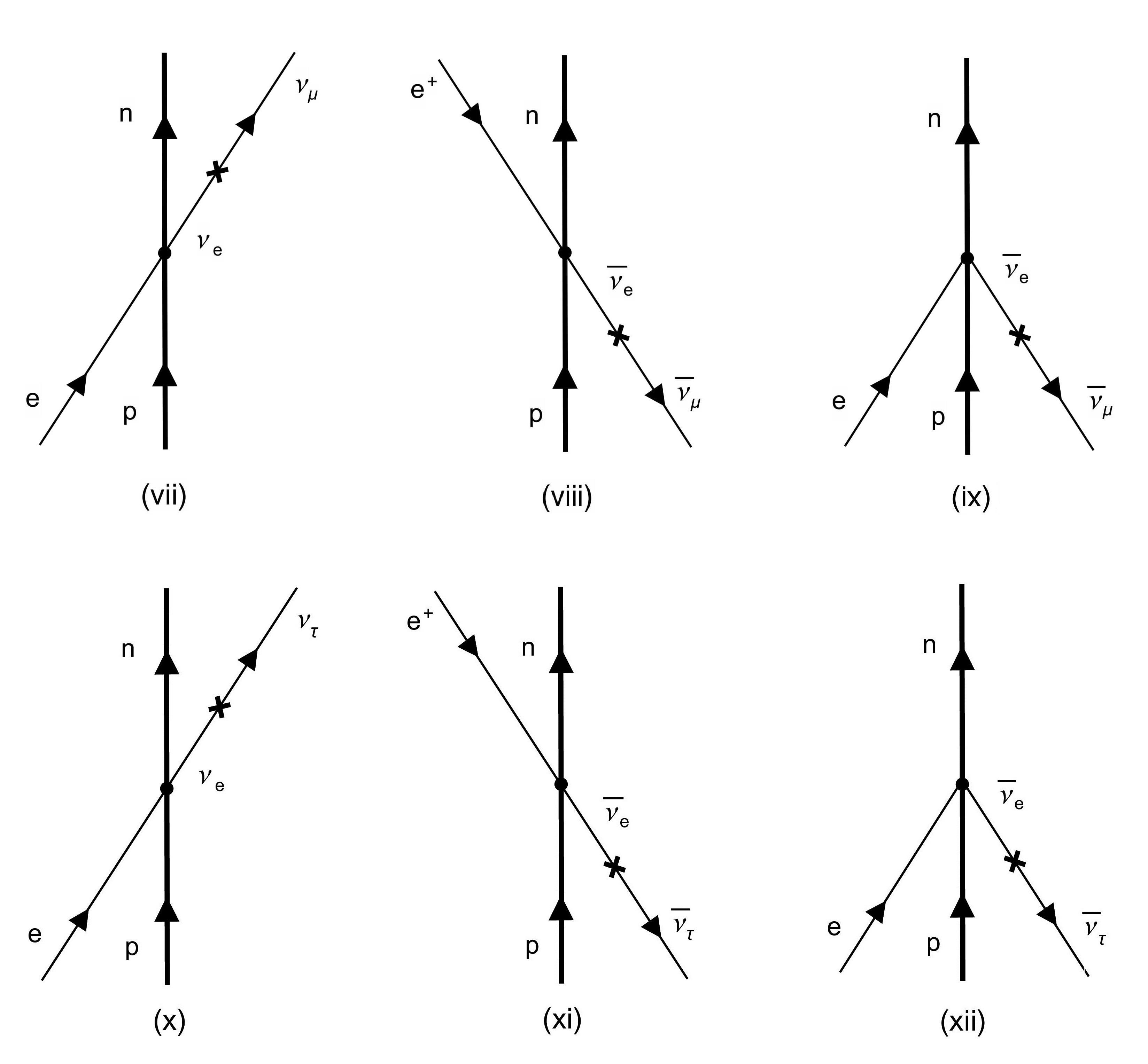}}}
\caption{Inverse $\beta$-decay in the comoving frame in the presence of flavor oscillations (time flows in the vertical direction).  The crosses denote the occurrence of the oscillation.}
\label{accoscillations}
\end{figure*}

In Sec.~\ref{LF} we have seen that, 
in addition to the flavor-conserving process~\eqref{lab},
the decay channels~\eqref{lab2} and \eqref{lab3} 
also have a non-vanishing probability because of 
the asymptotic occurrence of neutrino oscillations.
Guided by the principle of General Covariance of QFT, 
we thus search for the corresponding processes
to be considered in the comoving frame.
To this aim, we propose the following interactions
as candidates for the non-inertial counterparts of 
$\mathrm{(ii)}$ and $\mathrm{(iii)}$ (see Fig.~\ref{accoscillations})
\begin{subequations}
\label{comovoscu}
\begin{align}
&\mathrm{(vii)}\quad p\,+\,e^{-}\,\rightarrow\, n\,+\,\nu_\mu\\[2mm]
&\mathrm{(viii)}\quad p\,+\,\overline{\nu}_\mu\,\rightarrow\, n\,+\,e^{+}\\[2mm]
&\mathrm{(ix)}\quad p\,+\,e^{-}\,+\,\overline{\nu}_\mu\,\rightarrow\, n\,,
\end{align}
\end{subequations}
and
\begin{subequations}
\label{comovoscd}
\begin{align}
&\mathrm{(x)}\quad p\,+\,e^{-}\,\rightarrow\, n\,+\,\nu_\tau\,,\\[2mm]
&\mathrm{(xi)}\quad p\,+\,\overline{\nu}_\tau\,\rightarrow\, n\,+\,e^{+}\,,\\[2mm]
&\mathrm{(xii)}\quad p\,+\,e^{-}\,+\,\overline{\nu}_\tau\,\rightarrow\, n\,.
\end{align}
\end{subequations}
Note that, whilst the processes $\mathrm{(vii)}$ and $\mathrm{(x)}$
are of the same type as $\mathrm{(ii)}$ and $\mathrm{(iii)}$
in Eqs.~\eqref{lab2} and \eqref{lab3}, since 
they only provide for the oscillation of the emitted (electron)
neutrino, the remaining channels in Eqs.~\eqref{comovoscu}
and~\eqref{comovoscd} bring new physics into play 
with respect to $\mathrm{(v)}$ and $\mathrm{(vi)}$ in~\eqref{comov}, 
as they require that a muon- or tau-antineutrino 
in the Unruh thermal bath oscillates into an
electron-antineutrino before being absorbed by the proton (we remind that, 
at tree-level, the lepton charge must be conserved in the 
interaction vertices).

For the above processes, the decay rates can be
evaluated in the same way as in Eq.~\eqref{first}, yielding 
\begin{eqnarray}
\non
\label{TIR2}
\Gamma^{(\mathrm{vii})}\,+\,\Gamma^{(\mathrm{viii})}\,+\,\Gamma^{(\mathrm{ix})}&=&{|U_{\mu1}|}^2{|U_{e1}|}^2\,\,\widetilde\Gamma_1\,+\,{|U_{\mu2}|}^2{|U_{e2}|}^2\,\,\widetilde\Gamma_2\\[2mm]\non
&&\hspace{-34mm}+\,{|U_{\mu3}|}^2{|U_{e3}|}^2\,\,\widetilde\Gamma_3\,+\,\Big(U_{\mu1}^*\,U_{e1}\,{U_{\mu2}\,U^*_{e2}}\,\,\widetilde\Gamma_{12}\\[2mm]
&&\hspace{-34mm}+\,U_{\mu1}^*\,U_{e1}\,{U_{\mu3}\,U^*_{e3}}\,\,\widetilde\Gamma_{13}\,+\,U_{\mu2}^*\,U_{e2}\,{U_{\mu3}\,U^*_{e3}}\,\,\widetilde\Gamma_{23}\,+\,\mathrm{c.c.}\Big),
\end{eqnarray}
\vspace{1mm}
\begin{eqnarray}
\non
\label{TIR3}
\Gamma^{(\mathrm{x})}\,+\,\Gamma^{(\mathrm{xi})}\,+\,\Gamma^{(\mathrm{xii})}&=&{|U_{\tau1}|}^2{|U_{e1}|}^2\,\,\widetilde\Gamma_1\,+\,{|U_{\tau2}|}^2{|U_{e2}|}^2\,\,\widetilde\Gamma_2\\[2mm]\non
&&\hspace{-31.5mm}+\,{|U_{\tau3}|}^2{|U_{e3}|}^2\,\,\widetilde\Gamma_3\,+\,\Big(U_{\tau1}^*\,U_{e1}\,{U_{\tau2}\,U^*_{e2}}\,\,\widetilde\Gamma_{12}\\[2mm]
&&\hspace{-31.5mm}+\,U_{\tau1}^*\,U_{e1}\,{U_{\tau3}\,U^*_{e3}}\,\,\widetilde\Gamma_{13}\,+\,U_{\tau2}^*\,U_{e2}\,{U_{\tau3}\,U^*_{e3}}\,\,\widetilde\Gamma_{23}\,+\,\mathrm{c.c.}\Big),
\end{eqnarray}
where $\widetilde\Gamma_j$ and $\widetilde\Gamma_{jk}$
have been defined in Eqs.~\eqref{integralbis} and \eqref{int1212}, respectively. 
As a consequence, the total decay rate in the comoving system takes the form\begin{equation}
\Gamma^{\mathrm{com}}\,\equiv\,\sum_{s\hspace{0.2mm}=\hspace{0.2mm}\mathrm{iv}}^{\mathrm{xii}}\Gamma^{{(s)}}\,=\,{|U_{e1}|}^2\,\,\widetilde\Gamma_1\,+\,{|U_{e2}|}^2\,\,\widetilde\Gamma_2\,+\,{|U_{e3}|}^2\,\,\widetilde\Gamma_3\,.
\label{comtot}
\end{equation}
Now one can prove that the following equalities hold:
\begin{equation}
\label{seceq1}
\Gamma_{j}=\widetilde\Gamma_{j}\,,
\ee
and
\be
\Gamma_{jk}=\widetilde\Gamma_{jk}\,,
\label{seceq}
\ee
with Eq.~\eqref{seceq} being valid at least in the approximation
of small mass differences\footnote{Formally, one can prove each of the three equalities in Eq.~\eqref{seceq} by employing the same assumptions as in Ref.~\cite{NOUR}.}. 
Hence, by use of the above relations, it follows that the decay rates
for each neutrino flavor in the laboratory and 
comoving frames are in agreement with each other, i.e. 
\begin{eqnarray}
\label{49}
\Gamma^{(\mathrm{i})}&=&\Gamma^{(\mathrm{iv})}\,+\,\Gamma^{(\mathrm{v})}\,+\,\Gamma^{(\mathrm{vi})}\,,\\[2mm]
\label{50}
\Gamma^{(\mathrm{ii})}&=&\Gamma^{(\mathrm{vii})}\,+\,\Gamma^{(\mathrm{viii})}\,+\,\Gamma^{(\mathrm{ix})}\,,\\[2mm]
\label{51}
\Gamma^{(\mathrm{iii})}&=&\Gamma^{(\mathrm{x})}\,+\,\Gamma^{(\mathrm{xi})}\,+\,\Gamma^{(\mathrm{xii})}\,.
\end{eqnarray}
Equations~\eqref{49}-\eqref{51} naturally imply that
\be
\Gamma^{\mathrm{lab}}\,=\,\Gamma^{\mathrm{com}}\,.
\ee
This substantiates the result
that neutrino mixing 
is consistent with the General Covariance 
of QFT, since the (scalar) decay rate is independent
of the reference frame. In passing, we mention that
the opposite outcome is exhibited in Ref.~\cite{Ahluw}, 
where the equality between the two rates 
is claimed to be spoilt when taking into account flavor mixing. 
From comparison with our framework, it is clear that 
such a contradiction originates from the fact that the authors of 
Ref.~\cite{Ahluw} assume flavor neutrinos as fundamental objects in the laboratory frame, 
while they choose the mass representation in the comoving system.
A further argument for flavor states will be given in the
next Section, where we highlight the inadequacy 
of the mass representation to describe  
CP asymmetry in neutrino oscillations.

\section{CP violation in neutrino oscillations in Unruh radiation}
\label{Inconst}
It is well-known that
neutrino oscillations in the three-flavor description
can exhibit 
non-trivial CP-violation effects~\cite{Dick:1999ed, Nunokawa, CP}. 
Quantitatively speaking, the size of these effects
is controlled by the Jarlskog invariant $J$, 
which is a phase-convention-independent 
measure of CP violation in the Standard Model.
Despite being originally introduced in the context of 
quark mixing~\cite{Jar}, the definition of the Jarlskog invariant
can be straightforwardly rephrased in terms of 
the PMNS matrix~\eqref{PMM} for neutrinos as follows
\be
\label{Ji}
\mathrm{Im}\left[{U_{\delta i}}U^*_{\gamma_i}U^*_{\delta j}U_{\gamma j}\right]\,\equiv\, J\sum_{\lambda,k}\varepsilon_{\delta \gamma \lambda}\hspace{0.4mm}\varepsilon_{i j k}\,,
\ee
where $\delta,\gamma,\lambda=\{e,\mu,\tau\}$ and $i,j,k=\{1,2,3\}$. 
By using the parameterization~\eqref{PMM} for the PMNS matrix, $J$
can be explicitly written as
\begin{eqnarray}\non
J&=&c_{12}\hspace{0.2mm}s_{12}\hspace{0.2mm}c_{23}\hspace{0.2mm}s_{23}\hspace{0.2mm}c^2_{13}\hspace{0.2mm}s_{13}\sin\delta\\[2mm]
&=&\frac{1}{8}\sin2\theta_{12}\hspace{0.2mm}\sin2\theta_{23}\hspace{0.2mm}\cos\theta_{13}\hspace{0.2mm}\sin2\theta_{13}\hspace{0.2mm}\sin\delta\,.
\end{eqnarray}
Clearly, since physical quantities cannot depend on the choice of the
parameterization of the PMNS matrix, all CP-violating 
observables must depend on the invariant $J$ solely~\cite{Jar}. 
In passing, we mention that
a useful way of representing CP violation 
are the unitarity triangles (see Fig.~\ref{triangle}). These are constructed
exploiting the unitarity of the matrix~\eqref{PMM}, which
implies that different rows or columns are orthogonal to each other. 
For instance, we have
\be
\sum_{j=1}^3U_{\ell j}U^*_{\ell' j}\,=\,0\,, \quad \ell\neq\ell'\,.
\ee
The above relation can be represented as
a unitarity triangle in the complex plane by 
drawing arrows corresponding to the numbers
$U_{\ell j}U^*_{\ell' j}$ etc., and arranging the tip of each
arrow in such a way it coincides with the base of another 
(the orientation of these triangles 
has no physical meaning since, under
rephasing transformations, they simply rotate in the
complex plane). One can construct different
unitarity triangles, depending on which row or column
is considered. In spite of this, their area is 
invariant, being one-half the Jarlskog invariant introduced
in Eq.~\eqref{Ji}. Hence, it provides a measure
of CP violation. It goes without saying that, if all the elements of the PMNS matrix 
are real (i.e. if there is not CP asymmetry), the unitarity triangles
collapse in a line of vanishing area, as expected from the condition
$J=0$.

Starting from the above considerations, 
let $S_{weak}$ be the scattering matrix of a 
given charged-current weak interaction. 
In order to study CP-violation 
effects in a neutrino flavor-changing process, we assume
that a neutrino of a certain flavor (e.g. an electron neutrino)
is produced in the final state, so that $S_{weak}$ will depend on 
Dirac bilinears containing the (electron) neutrino field $\psi_{\nu_e}$ 
(we may refer, for example, 
to the inverse $\beta$-decay discussed above, as well as to
any other similar process involving a neutrino in the final state).
By describing
asymptotic neutrinos by means of flavor states, 
the probability
that, after being emitted, the neutrino is detected 
with a different flavor (for instance, as a muon neutrino) can be derived
from the following transition amplitude:
\be
\label{Amp}
\mathcal{A}_{\nu_e,\nu_\mu}\,=\,_{out}\langle \nu_\mu,\dots| S_{weak}\big(\bar\psi_{\nu_e}\dots\big)|\dots\rangle_{in}\,,
\ee
where the first (second) subscript in the l.h.s.
refers to the neutrino field (state) appearing in the $S$-matrix
in the r.h.s., and the dots must be filled with the appropriate 
fields and particles involved in the considered interaction.

\begin{figure}[t]
\centering
{\resizebox{8.5cm}{!}{\includegraphics{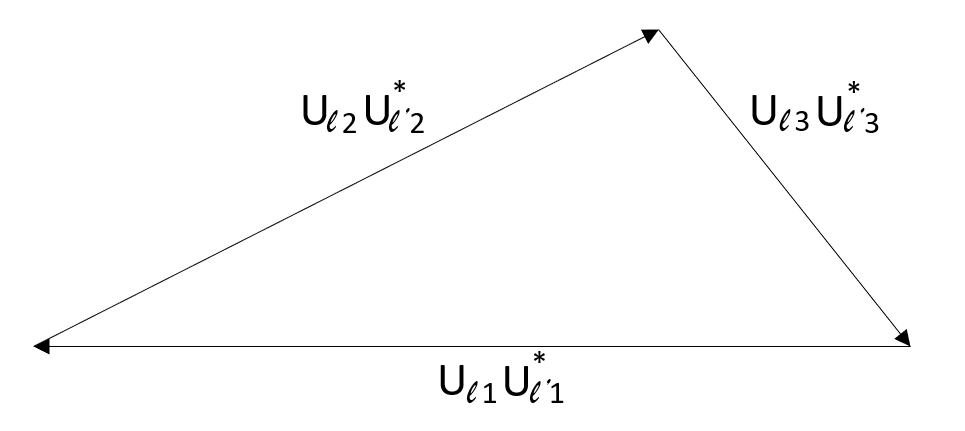}}}
\caption{Schematic representation of a unitarity triangle.}
\label{triangle}
\end{figure}

If we now implement the mixing transformations~\eqref{eqn:U}
and \eqref{field} on neutrino state and field, Eq.~\eqref{Amp} can be cast in the form
\be
\label{Amp1}
\mathcal{A}_{\nu_e,\nu_\mu}\,=\,\sum_{j=1}^3 U^*_{\mu j}\hspace{0.2mm}U_{ej}\,\mathcal{A}_{j}\,,
\ee
where we have used the shorthand notation
\be
\mathcal{A}_{j}\,\equiv\,_{out}\langle \nu_j,\dots| S_{weak}\big(\bar\psi_{\nu_j}\dots\big)|\dots\rangle_{in}\,.
\ee
Thus, by exploiting the definition~\eqref{scaldeca}
and squaring both sides of Eq.~\eqref{Amp1}, we obtain 
\begin{eqnarray}
\label{appdr}
\non
\Gamma_{\nu_e,\nu_\mu}&\sim&{|\mathcal{A}_{\nu_e,\nu_\mu}|}^2=\\[2mm]\non
&&\hspace{-13mm}{|U_{\mu1}|}^2{|U_{e1}|}^2{|\mathcal{A}_{1}|}^2+\,{|U_{\mu2}|}^2{|U_{e2}|}^2{|\mathcal{A}_{2}|}^2\,+\,{|U_{\mu3}|}^2{|U_{e3}|}^2{|\mathcal{A}_{3}|}^2\,\\[2.5mm]\non
&&\hspace{-11mm}+\Big(U_{\mu1}^*\,U_{e1}\,{U_{\mu2}\,U^*_{e2}}\hspace{0.6mm}\mathcal{A}_{1}\,\mathcal{A}^*_{2}+\,U_{\mu1}^*\,U_{e1}\,{U_{\mu3}\,U^*_{e3}}\hspace{0.6mm}\mathcal{A}_{1}\,\mathcal{A}^*_{3}\\[2mm]
&&\hspace{-11mm}+\,\,U_{\mu2}^*\,U_{e2}\,{U_{\mu3}\,U^*_{e3}}\hspace{0.6mm}\mathcal{A}_{2}\,\mathcal{A}^*_{3}\,+\,\mathrm{c.c.}\Big).
\end{eqnarray}
where, in order not to burden the notation, we have omitted 
the sum over polarizations and the integration over momenta
of the emitted particles in the r.h.s.

To quantify CP asymmetry in the above interaction, we now assume that
particles swap places with their antiparticles while viewed in a mirror. By computing 
the decay rate $\Gamma_{\bar\nu_e,\bar\nu_\mu}$ for
the ensuing process, we finally arrive at 
\begin{eqnarray}
A^{(e,\mu)}_{CP}&\equiv&\Gamma_{\nu_e,\nu_\mu}\,-\,\Gamma_{\bar\nu_e,\bar\nu_\mu}\\[2mm]
\non
&=&4\hspace{0.2mm}J \Big\{-\mathrm{Im}\left[\mathcal{A}_1\mathcal{A}^*_2\right]\,+\,\mathrm{Im}\left[\mathcal{A}_1\mathcal{A}^*_3\right]\,-\,\mathrm{Im}\left[\mathcal{A}_2\mathcal{A}^*_3\right]\Big\},
\end{eqnarray}
which is indeed non-vanishing and does not depend on the specific parameterization of the mixing matrix, as it should be.
It is worth noting that, if we calculate the same quantity for
the transition between the electron- and tau-neutrino flavors, from Eq.~\eqref{Ji} 
we obtain
\begin{eqnarray}
A^{(e,\tau)}_{CP}&\equiv&\Gamma_{\nu_e,\nu_\tau}\,-\,\Gamma_{\bar\nu_e,\bar\nu_\tau}\\[2mm]\non
&=&4\hspace{0.2mm}J \Big\{\mathrm{Im}\left[\mathcal{A}_1\mathcal{A}^*_2\right]\,-\,\mathrm{Im}\left[\mathcal{A}_1\mathcal{A}^*_3\right]\,+\,\mathrm{Im}\left[\mathcal{A}_2\mathcal{A}^*_3\right]\Big\}.
\end{eqnarray}
By adding up $A^{(e,\mu)}_{CP}$ and $A^{(e,\tau)}_{CP}$, 
it follows that
\be
A^{(e,e)}_{CP}\,+\,A^{(e,\mu)}_{CP}\,+\,A^{(e,\tau)}_{CP}\,=\,0\,,
\ee
where we have exploited the fact that
a CP asymmetry can be measured only in transitions
between different flavors, i.e. $A^{(e,e)}_{CP}=0$.

On the other hand, if one adopts the point of view of Ref.~\cite{Cozzella} 
and assumes the mass representation as the fundamental one, 
the above CP-violation feature does not emerge at all. 
Indeed, by straightforward calculations, one has
\begin{eqnarray}
\label{ACP}
A_{CP}^{(e,j)}&=&\Gamma_{\nu_e,\nu_j}\,-\,\Gamma_{\bar\nu_e,\bar\nu_j}\\[2mm]\non
&=&{|U_{ej}|}^2\hspace{0.2mm}{|\mathcal{A}_j|}^2\,-\,{|U^*_{ej}|}^2\hspace{0.2mm}{|\mathcal{A}_j|}^2\,=\,0\,,\,\,\quad j=1,2,3\,,
\end{eqnarray}
which clearly shows that mass states are inconsistent with
the picture of CP violation in the neutrino sector. 
At the same time, however, it is immediate to see that
\be
\sum_{\ell=e,\mu,\tau} A^{(e,\ell)}_{CP}\,=\,\sum_{j=1}^{3} A^{(e,j)}_{CP}\,=\,0\,,
\ee
as it might be expected from Eq.~\eqref{totdecrat} and the related discussion.

Finally, with reference to the inverse $\beta$-decay
analyzed above and, in particular, to the processes 
$\mathrm{(viii)}$, $\mathrm{(ix)}$, $\mathrm{(xi)}$ and $\mathrm{(xii)}$
in the comoving frame, we note that
the occurrence of CP violation
manifests itself in an asymmetry 
between the thermal baths 
experienced by the accelerated proton and antiproton, respectively: this originates from the different oscillating behavior between neutrinos and anti-neutrinos in the Unruh radiation. Clearly, this is a novel feature 
of the Unruh effect which
does not appear in the previous 
literature on the inverse $\beta$-decay 
with mixed neutrinos~\cite{NMIBD,NOUR,Cozzella}, 
since it is peculiar of the three-flavor description.

\section{Discussion and Conclusions}
\label{DeC}
The inverse $\beta$-decay of uniformly accelerated protons
has been investigated in the context of three-flavor neutrino mixing
and oscillations. By assuming neutrinos to be Dirac particles
and working within the $S$-matrix framework, we have shown
that the decay rates in the laboratory and comoving
frames agree with each other, provided that the asymptotic behavior of 
neutrinos is described by means of flavor (rather than mass) eigenstates.
It is worth noting that such an analysis would be 
a rather straightforward generalization of the two-flavor 
treatment of Ref.~\cite{NOUR}, if it were not for the
presence of the Dirac phase in the PMNS matrix. As well known, such phase 
induces non-trivial CP-violation effects in neutrino oscillations, which determine
 an asymmetry between
the Unruh radiation detected 
by the accelerated proton and antiproton, respectively.  
In this connection, we have
proved that the mass representation 
is inconsistent with the 
picture of CP asymmetry in the neutrino sector, 
as it leads to
an identically vanishing expression for the quantity 
$A_{CP}$  (see Eq.~\eqref{ACP}). On the other hand, 
by adopting flavor asymptotic states, $A_{CP}$ turns out to be 
proportional to the Jarlskog rephasing-invariant, 
as one would expect for any physical observable which quantifies 
CP-violation. 

Despite the obtained equality between
the inertial and comoving decay rates, we emphasize that the analysis 
carried out here may not represent the end
of the story, since it holds in the approximation
of small neutrino mass differences (see the discussion
after Eq.~\eqref{seceq}). 
The question inevitably arises as to how accommodate 
next-to-leading order corrections without affecting the
internal consistency of the formalism. In this regard, we envisage that 
some new features might come into play, as for example the necessity of a full-fledged QFT treatment of neutrino mixing instead 
of the Pontecorvo quantum mechanical one~\cite{BV95}, 
or the possibility to violate the thermality of the Unruh effect 
when considering mixed fields~\cite{BLUC}. Note that 
similar non-thermal distortions of the Unruh-Hawking
spectrum have been recently highlighted also in 
other contexts, such as the emission on the background of a quantum collapsing null shell~\cite{CQS}, the polymer (loop) quantization for 
the calculation of the two-point function along Rindler trajectories~\cite{PLQ}, the
Casimir effect between uniformly accelerated atoms~\cite{Marino} and the Generalized Uncertainty Principle framework~\cite{Scard}. 

As remarked above, in our study we have considered the 
case of Dirac neutrinos. However, the question
about the very nature of neutrinos - Dirac or Majorana - is still open.
As it is well-known~\cite{Bilenky:1980cx}, oscillation experiments do not allow us
to discriminate between these two alternatives, 
the only feasible test being the neutrinoless double $\beta$-decay~\cite{Avignone}.
Also from the point of view of the Unruh effect, 
it has been shown that there is no difference
between the employment of Dirac and Majorana fermion fields
in the computation of the (accelerated) thermal distribution~\cite{LonghiSold}.
Thus, in light of the above considerations, 
we expect the overall validity of our analysis to be
unaffected by the nature of neutrinos, although some formal differences
may arise when working with Majorana fields, due to the presence of two additional phases in the mixing matrix. 
In particular, concerning General Covariance, 
we envisage that the equality between the decay rates in the two frames
must hold true, owing to the fact that the mixing matrix is still unitary. 
Likewise, one can repeat the same reasoning on CP violation 
as in Sec.~\ref{Inconst} and come up with the same conclusion, 
since the Majorana phases do not contribute to the Jarlskog invariant~\cite{Giuntibook}.

Apart from its intrinsic theoretical interest, we remark
that a deeper understanding of the very nature of asymptotic neutrino states
may also be relevant from the experimental point of view. 
In Ref.~\cite{ChengYang}, indeed, 
it has been shown that the spectrum of the Tritium $\beta$-decay 
near the end point energy is sensitive to whether 
neutrinos interact as massive or flavor eigenstates.
Similar considerations are valid
for the neutrino capture by Tritium too. 
In light of this, it is reasonable to expect
that accurate measurements from
such current experiments as KATRIN (which aims to appoint an upper limit to the electron antineutrino mass by examining 
the spectrum of electrons emitted from the Tritium $\beta$-decay)~\cite{Osipowicz:2001sq} and 
PTOLEMY (that is
projected to detect the cosmic neutrino background)~\cite{Pto}  
might provide important pieces of 
information in the problem at hand.

Finally, we highlight that the above study
is closely related to the issue of non-inertial/gravitational effects
on the oscillation probability. 
A preliminary investigation of this problem has been 
proposed for the case of accelerated 
systems~\cite{NOAF}, in 
curved spacetime~\cite{NOCB,NOCB2}, in 
astrophysical and cosmological regimes~\cite{CosmAstr,CosmAstr2}, in extended theories of gravity~\cite{Capoz} 
and in stochastic model for spacetime foam~\cite{foam}. 
Worthy of attention may be also quantum-gravity decoherence effects
in oscillations~\cite{mavr} and the entanglement 
among neutrinos and the other particles~\cite{EPLahluw} in decay processes. 
All of these issues are currently under active considerations.

\end{document}